# Flow regimes and types of solid obstacle surface roughness in turbulent heat transfer inside periodic porous media

**Vishal Srikanth[1], Dylan Peverall[1], and Andrey V. Kuznetsov[1,a]**

**Abstract**
The role of solid obstacle surface roughness in turbulent convection in porous media is not well understood, even though it is frequently used for heat transfer enhancement in many applications. The focus of this paper is to systematically study the influence of solid obstacle surface roughness in porous media on the microscale flow physics and report its effect on macroscale drag and Nusselt number. The Reynolds averaged flow field is numerically simulated using the realizable $k$-$\varepsilon$ model for a flow through a periodic porous medium consisting of an in-line arrangement of square cylinders with square roughness particles on the cylinder surface. Two flow regimes are identified with respect to the surface roughness particle height – fine and coarse roughness regimes. The effect of the roughness particles in the fine roughness regime is limited to the near-wall boundary layer around the solid obstacle surface. In the coarse roughness regime, the roughness particles modify the microscale flow field in the entire pore space of the porous medium. In the fine roughness regime, the heat transfer from the rough solid obstacles to the fluid inside the porous medium is less than that from a smooth solid obstacle. In the coarse roughness regime, there is an enhancement in the heat transfer from the rough solid obstacle to the fluid inside the porous medium. Total drag reduction is also observed in the fine roughness regime for the smallest roughness particle height. The surface roughness particle spacing determines the fractional area of the solid obstacle surface covered by recirculating, reattached, and stagnating flow. As the roughness particle spacing increases, there are two competing factors for the heat transfer rate – increase due to more surface area covered by reattached flow, and decrease due to the decrease in the number of roughness particles on the solid obstacle surface. Decreasing the porosity and increasing the Reynolds number amplify the effect of the surface roughness on the microscale flow. The results suggest that heat transfer in porous media can be enhanced, if the increase in drag can be overcome. The results also show that the fine roughness regime, which is frequently encountered due to corrosion, is detrimental to the heat transfer performance of porous media.

## 1   Introduction

Surface roughness is commonly encountered in both natural and engineered porous media. Roughness can be formed at the time of manufacturing or develop over time due to corrosion, scale formation, or wear and tear. Since surface roughness modifies the solid obstacle geometry, it influences the microscale turbulent flow around the solid obstacle. The influence of the surface roughness on the macroscale flow is compounded by the presence of numerous solid obstacles in porous media. Therefore, it is a critical parameter in the modeling of turbulent flow and convection heat transfer in porous media. There are few studies on the effect of solid obstacle surface roughness on transport in porous media, and a majority of the studies focus on laminar multiphase flow. The studies consider applications to specific types of porous media such as soil and porous rocks that have small void spaces due to low porosity. In these scenarios, there is evidence of retardation in multiphase flows even for small roughness heights based on the experiments of

[a]Email address for correspondence: avkuznet@ncsu.edu
[1]Department of Mechanical and Aerospace Engineering, North Carolina State University, Raleigh, NC 27695, USA

Mehmani et al. (2019), Lyu et al. (2020), and Zhang et al. (2021). This retardation occurs due to the drag from the roughness particles on the fluids as well as the interaction of the two-phase fluid interface with the roughness particles.

There are a few analytical and numerical models studying the effect of surface roughness on single phase flow. The single-phase permeability of a porous medium with rough surfaces is dependent on the size of the pores and the roughness particles, as per the analytical model of Yang et al. (2015). The lattice Boltzmann simulations of 3D fractal models of laminar flow through a porous medium with rough solid obstacles confirm that the permeability decreases due to the introduction of surface roughness (Cousins et al. 2018). Pasquier et al. (2017) performed numerical studies of laminar flows past a porous medium consisting of an array of circular cylinders with sinusoidal roughness grooves. They note that the roughness introduces inertial effects and viscous dissipation at large Reynolds numbers within the laminar flow regime. For in-line tube banks, surface roughness on the tube surface was shown to improve heat transfer (Achenbach 1991). In the experiment described in Achenbach (1991), pyramid-shaped roughness was created by knurling the tube surface with roughness heights that are 100 times less than the tube diameter. The heat transfer enhancement was accompanied by the increase in the pressure drop across the tube bank to keep the Reynolds number the same. To the best of the authors' knowledge, there are no other studies on the effect of solid obstacle surface roughness on turbulent heat transfer in porous media.

There are several studies on the effect of surface roughness on flows around bluff bodies and inside channels, which can provide a baseline for our analysis. For turbulent channel flows, the roughness particles increase the mean Reynolds shear stresses, even in the outer boundary layer (Choi et al. 2020; Tay et al. 2013). The influence of the roughness particles close to the rough wall is dependent on the roughness particle geometry. For rib type roughness (similar to the kind used in the present work) and ratchet type roughness, the roughness particles introduced their own characteristic length scales to the near wall turbulence structures (Busse and Zhdanov 2022; Cui et al. 2003). The geometry of the roughness particles modifies the properties of the flow structures and the boundary layer at the wall. Jin and Herwig (2014) demonstrate the use of shark fin inspired roughness for drag reduction due to the turbulent structure lift-off from the wall caused by the roughness particles. In contrast, a roughness model for conical roughness particles predicts additional drag and turbulence dissipation at the roughness layer (Miyake et al. 2000). Heat transfer is also influenced by the roughness particle geometry due to its effect on momentum transport in the roughness layer. MacDonald et al. (2019) show that heat transfer continues at the rough wall with wavy roughness through a thin heat diffusion layer even though the velocity in this layer is substantially decreased. Depending on the roughness particle height and spacing, heat transfer enhancement is experienced along with an increase in the drag (Katoh et al. 2000; Nagano et al. 2004).

The flow behavior around a rough circular cylinder exhibits some similarities when compared to that of a rough channel. Since there are no streamwise periodic boundaries in the flow around a circular cylinder, roughness on the circular cylinder accelerates the transition to turbulence near the surface (Rodriguez et al. 2016). Transition to turbulence is not expected in the present work for porous media, so the roughness particles will only modify the already turbulent boundary layer. The shear layer around a rough cylinder surface is disturbed by the roughness particles (Sun et al. 2022). As a result, the roughness particles change the location of the separation points on the surface in circular cylinders (Jiang et al. 2017). Early flow separation on the cylinder surface increases the size of the flow recirculation area, which in turn increases the drag (Yamagishi and Oki 2007). The shape of the roughness particle determines the magnitude of the increase in the size of the flow recirculation area (Yamagishi and Oki 2004). On the other hand, Zhou et al. (2015) show that rough cylinders can result in a compact wake that results in reduced drag. Although not explicitly stated, the roughness particle height and shape determine whether drag is increased or reduced by the roughness particles. Dierich and Nikrityuk (2013) analyzed the heat transfer from rough circular cylinders and concluded that the increase in the roughness height decreases the Nusselt number. It should be noted that the increase in the roughness height in this study caused the formation of deep trenches on the cylinder surface that are isolated from the flow around the cylinder.

Previous literature reports that roughness can cause both heat transfer enhancement and reduction, as well as drag increase and reduction. This is because the roughness particle geometries used in these studies are different. There is a strong dependence of the flow around rough solid obstacles on the geometry of the roughness particles. A systematic study on the microscale flow physics that is caused by the roughness particles is lacking, which the present work addresses. The objective of the present work is to study the effect of the surface roughness particles on turbulent convection heat transfer in porous media. Porous media combine the features of flow inside channels and flow around bluff bodies. Additional considerations such as the porosity and the Reynolds number increase the complexity of the analysis of the microscale flow. We will perform the systematic study by incrementally varying the surface roughness particle height and spacing while considering different flow regimes with respect to porosity and the Reynolds number. To limit the scope of the paper, the roughness particle shape, aspect ratio, and three dimensionality are not explored, however they are important parameters for future work. The Reynolds averaged flow solution will be used to determine the trends in the macroscale quantities, such as the pressure drag, the viscous drag, and the surface averaged Nusselt number, with the surface roughness length scales of roughness particle height and spacing. We will emphasize the underlying flow physics that causes the macroscale trends by analyzing the microscale flow patterns that are caused by the roughness particles. We define flow regimes and types of roughness based on the underlying flow physics so that they are classified into fundamental flow situations. The details of the geometry used to simulate the porous medium are described in section 2.1. The numerical procedure is described in section 2.2, followed by a validation study in section 2.3. The results are discussed in section 3.

## 2 Solution Method

## 2.1 Computational geometry

The focus of the present work is to determine the influence of solid obstacle surface roughness on flow in porous media at the microscale and the effect that it has on macroscale parameters. To study this, an infinitely spanning porous medium is represented by an in-line arrangement of square cylinder solid obstacles. The use of the Reynolds Averaged Navier-Stokes (RANS) approach allows for a two-dimensional (2D) approximation of the porous medium geometry since the flow variables are not expected to vary in the axial direction of the square cylinder. Note that if the governing equation in the axial direction of the cylinder is solved, the velocity and the gradients of the independent variables in the axial direction of the cylinder will yield a trivial result equal to zero. The RANS approach also permits the use of a Representative Elementary Volume (REV) consisting of only one solid obstacle due to the following reasoning. Since the scales of turbulence are not resolved in the RANS approach, they are not a consideration for selecting the size of the computational domain. The solution of the RANS governing equations will periodically repeat across the unit cells in the infinitely spanning porous medium due to the geometry invariance. After the Reynolds averaging operation, the turbulence statistics surrounding the solid obstacles is identical for all of the solid obstacles in the periodic porous medium. Simulating an REV that is the size of a single unit cell with the use of periodic boundary conditions is adequate to determine the Reynolds averaged flow solution (Fig. 1).

The hydraulic diameter of the smooth solid obstacle, which is the side length of the square obstacle ($d$), is chosen as the characteristic length scale of the flow. To non-dimensionalize the length scales, the hydraulic diameter ($d$) is set equal to 1 m and the remaining length scales are reported as the ratio of the hydraulic diameter. Square roughness particles are placed on the surface of the square cylinder as shown in Fig. 1 to explicitly represent the solid obstacle surface roughness. The roughness particle height $k_s$ is equal to the side length of the square roughness particle. The distribution of the roughness particles is varied by changing the distance between the centroids of the roughness particles called the roughness particle spacing ($w$). The roughness particle spacing ($w$) controls the number of roughness particles ($n_p$) present along one side length of the square solid obstacle (equation 1). The values of $w$ in section 3 are chosen such that fractional roughness particles are not encountered. The porosity ($\varphi$) of the porous medium can be calculated using equation 2. Since the values of $d$, $k_s$, $w$, and $\varphi$ are prescribed in the cases shown in section 3, the value of $s$ changes for each case as per equation 2.

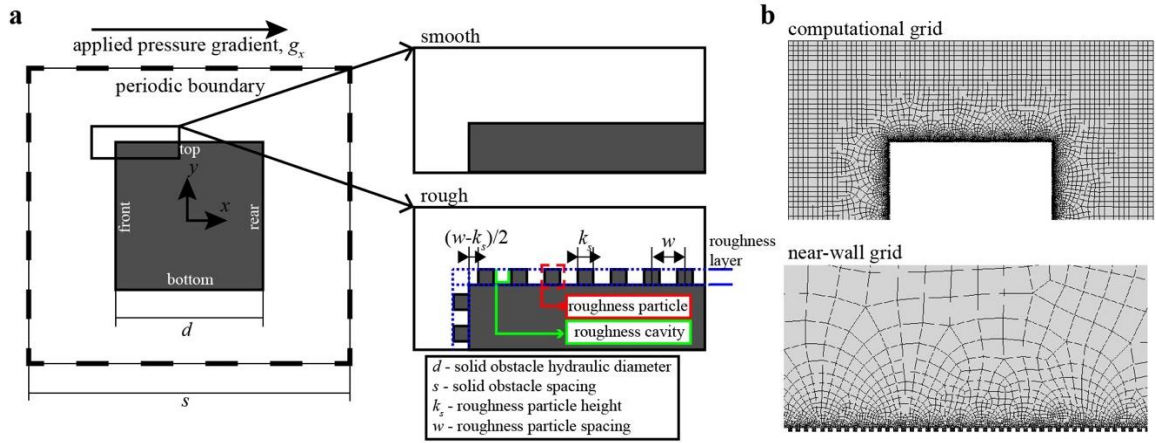

**Fig. 1** The computational geometry (**a**) and grid (**b**) used to simulate turbulent flow in periodic porous media with smooth and rough solid obstacle surfaces. The computational grid is shown for the case of $k_s = 0.005$, $w = 2k_s$, $\varphi = 0.8$ and it includes grid refinement at the solid obstacle wall.

$$n_p = \begin{cases} 0 & \text{; smooth} \\ d/w & \text{; rough} \end{cases} \tag{1}$$

$$\varphi = 1 - \frac{d^2 + 4n_p k_s^2}{s^2} \tag{2}$$

The microscale flow velocity and pressure distributions inside the REV are periodic in both the $x$ and $y$ directions. A constant flow rate is sustained inside the porous medium by applying a momentum source term that causes a decreasing pressure gradient in the flow direction ($g_x$). A microscale temperature gradient is imposed in the $x$ direction such that the bulk temperature of the fluid that enters the REV is 300 K. The solid obstacle surface has a constant temperature ($T_w$) of 350 K. Therefore, the fluid is heated by the solid obstacles as is flows downstream. The microscale temperature distribution in the $y$ direction in periodic. The computational grid is unstructured as shown in Fig. 1b. The number of computational grid cells ranges between 25,000 and 35,000 cells.

## 2.2 Details of the numerical method

In the present work, we are interested in determining the influence of rough solid obstacles on the Reynolds averaged flow inside the porous medium. The numerical simulations are performed using the commercial CFD solver Ansys® Academic Research Fluent, Release 17.0. The governing equations of the flow are the Reynolds Averaged Navier-Stokes (RANS) equations, which are solved numerically along with the realizable $k$-$\varepsilon$ model (Shih et al. 1995) to predict the turbulence viscosity. RANS simulations using linear $k$-$\varepsilon$ models have been used previously by researchers (Abed and Afgan 2017; de Lemos 2012; Pedras and de Lemos 2003) for porous media flows with good qualitative agreement of the flow patterns with LES (Iacovides et al. 2014). The realizable $k$-$\varepsilon$ model is more accurate in predicting flows involving rotation, boundary layers under strong adverse pressure gradients, separation, and recirculation when compared to the original $k$-$\varepsilon$ model by Launder and Spalding (1974). The two-layer $k$-$\varepsilon$, $k$-$l$ near-wall treatment (Chen and Patel 1988) is applied at the wall boundaries to circumvent the need for the specification of turbulence dissipation rate at the wall. Therefore, the simulation is wall resolved by including grid refinement at the solid obstacle surface.

The RANS equations with the linear eddy-viscosity approximation of the Reynolds stresses are written in equations 3-4 (the overbar notation denotes Reynolds averaging), where $\rho$ is the fluid density, $\mu$ is the dynamic viscosity, $u_i$ is the microscale velocity vector, $p$ is the periodic component of the static pressure, and $\rho g_i$ is the linearly declining component of the static pressure per unit length of the periodic domain. The realizable $k$-$\varepsilon$ model approximates the turbulence eddy viscosity ($\mu_T$) by solving equations 5-6 for the turbulence kinetic energy ($k$) and the turbulence dissipation rate ($\varepsilon$). The turbulence eddy viscosity is calculated using equation 8. The model is closed according to the constants and relations provided in the work of Shih et al. (1995).

$$\frac{\partial}{\partial x_i}(\rho \bar{u}_i) = 0 \tag{3}$$

$$\frac{\partial}{\partial x_j}\left(\rho \bar{u}_i \bar{u}_j\right) = -\frac{\partial \bar{p}}{\partial x_i} + \frac{\partial}{\partial x_j}\left[(\mu + \mu_T)\left(\frac{\partial \bar{u}_i}{\partial x_j} + \frac{\partial \bar{u}_j}{\partial x_i}\right)\right] + \rho g_i \tag{4}$$

$$\frac{\partial(\rho k \bar{u}_j)}{\partial x_j} = \frac{\partial}{\partial x_j}\left[\left(\mu + \frac{\mu_T}{\sigma_k}\right)\frac{\partial k}{\partial x_j}\right] + 2\mu_T \bar{S}_{ij}\bar{S}_{ij} - \rho \varepsilon \tag{5}$$

$$\frac{\partial(\rho \varepsilon \bar{u}_j)}{\partial x_j} = \frac{\partial}{\partial x_j}\left[\left(\mu + \frac{\mu_T}{\sigma_\varepsilon}\right)\frac{\partial \varepsilon}{\partial x_j}\right] + \rho C_1 \sqrt{2\bar{S}_{ij}\bar{S}_{ij}}\,\varepsilon - \rho C_2 \frac{\varepsilon^2}{k+\sqrt{\mu \varepsilon/\rho}} \tag{6}$$

where,

$$\bar{S}_{ij} = \frac{1}{2}\left(\frac{\partial \bar{u}_j}{\partial x_i} + \frac{\partial \bar{u}_i}{\partial x_j}\right) \tag{7}$$

$$\mu_T = \rho C_\mu \frac{k^2}{\varepsilon} \tag{8}$$

Near the wall, the transport equation for the turbulence dissipation rate is not solved. The value of $\varepsilon$ is estimated using equations 9-12. The one-equation model is invoked whenever the wall-distance-based Reynolds number (defined as $Re_y$ in equation 12; $y$ is the minimum distance to the solid wall) drops below a threshold value of 200, signifying the dominance of molecular viscosity.

A blending function is used to bridge the interface between the one- and two- equation models. The turbulence eddy viscosity is calculated using equation 13 near the wall for the one-equation model. The model constants are available in the ANSYS documentation (ANSYS Inc. 2016).

$$\varepsilon = \frac{k^{1.5}}{l_\varepsilon} \tag{9}$$

$$l_\varepsilon = y C_l^* \left( 1 - e^{\frac{-Re_y}{A_e}} \right) \tag{10}$$

$$l_\mu = y C_l^* \left( 1 - e^{\frac{-Re_y}{A_\mu}} \right) \tag{11}$$

$$Re_y = \frac{\rho y \sqrt{k}}{\mu} \tag{12}$$

Near the wall,

$$\mu_T = \rho C_\mu l_\mu \sqrt{k} \tag{13}$$

The governing equation for heat transfer, equation 14, is solved as a separate scalar equation for the conservation of energy, where $T$ is the microscale temperature distribution, $\lambda$ is the thermal conductivity (0.0242 W/m-K), and $C_P$ is the specific heat of the fluid (1006.43 J/Kg-K).

$$\frac{\partial(\bar{u}_i(\rho E + \bar{p}))}{\partial x_i} = \frac{\partial}{\partial x_j} \left( \lambda_{eff} \frac{\partial \bar{T}}{\partial x_j} + \bar{u}_i(\mu + \mu_T) \left( \frac{\partial \bar{u}_j}{\partial x_i} + \frac{\partial \bar{u}_i}{\partial x_j} - \frac{2}{3} \frac{\partial \bar{u}_k}{\partial x_k} \delta_{ij} \right) \right) \tag{14}$$

$$E = C_P(\bar{T} - 298.15) + 0.5 \bar{u}_j \bar{u}_j \tag{15}$$

$$\lambda_{eff} = \lambda + \frac{C_P \mu_T}{0.85} \tag{16}$$

The Finite Volume Method (FVM) is used to solve the governing equations. The derivatives are approximated using a second-order upwind scheme for the convective terms and a second-order central scheme for the viscous terms. The location of the pressure variable is staggered such that it is stored at the centroid of the face of the cell. The governing equations are solved in a segregated manner using the SIMPLE algorithm. The density is set equal to 1 and the dynamic viscosity is set equal to $1/Re$. The definition of the Reynolds number ($Re$) is provided in equation 17, where $u_m$ is the Reynolds averaged, superficially averaged $x$- velocity. The momentum source term $g_i$ is adjusted at every iteration to maintain a constant flow rate at the periodic boundaries such that $u_m$ is 1. Since the flow is incompressible, this translates to an indirect specification of the Reynolds number through the macroscale velocity. The values of $\rho$, $\mu$, $d$, and $u_m$ non-dimensionalize the governing equations of momentum. The Reynolds number is defined as:

$$Re = \frac{\rho u_m d}{\mu} \tag{17}$$

When the Reynolds number of the flow is changed in section 3, the viscosity of the fluid is changed accordingly as well as the specific heat to keep the Prandtl number at a constant value.

## 2.3 Validation of the numerical model

In this section, we demonstrate that the present numerical model is adequate to reproduce the flow features inside a porous medium. For comparison, we use the velocity and temperature profiles obtained from the direct numerical simulation (DNS) results reported by Chu et al. (2018) and Chu et al. (2019) for turbulent flow through a periodic porous medium composed of a staggered arrangement of smooth square cylinders. The porosity of the porous medium is 0.75. The Reynolds number of the flow is 1000 as per the definition in Chu et al. (2018) and 750 as per the definition used in this paper. The realizable $k$-$\varepsilon$ model used in the present work reproduces the following flow features that are observed in the DNS (Fig. 2b):

1. Flow separation and the formation of microscale vortices behind the solid obstacle.
2. Flow impingement and stagnation on the front of the solid obstacle.
3. Shear flow in between the horizontal surfaces of two neighboring solid obstacles.
4. Flow separation at the front vertices of the square geometry due to the tortuosity of the flow caused by the staggered obstacle arrangement.

The realizable $k$-$\varepsilon$ model assumes that the turbulence is isotropic, which is not valid in wall bounded flows such as in porous media. We cannot expect the realizable $k$-$\varepsilon$ model to accurately predict the detailed features of the velocity boundary layer that are present in the DNS results. Therefore, the velocity profile simulated by the realizable $k$-$\varepsilon$ model deviates the most from the DNS in the locations where the turbulence is anisotropic (Fig. 2a). However, the near-wall prediction of the velocity gradient of the realizable $k$-$\varepsilon$ model is reasonably close to that of the DNS. The near-wall velocity gradient is important to predict the viscous drag force that we have reported in section 3. We observe better agreement between the realizable $k$-$\varepsilon$ model and DNS for the non-dimensional temperature profile. The non-dimensional temperature is calculated as per the work of Patankar et al. (1977). The non-dimensional temperature profile predicted by the realizable $k$-$\varepsilon$ model deviates from that of the DNS at the locations of maximum deviation in the velocity profile. The prediction of the near-wall temperature gradient by the realizable $k$-$\varepsilon$ model is close to that of the DNS, which is important to predict the Nusselt number that we have reported in section 3.

We also determined the quality of the numerical method in each case by calculating the numerical error in satisfying the conservation of momentum. We evaluated the numerical error by considering the entire REV as the control volume for all of the cases reported in section 3. The maximum numerical error is 0.6% and the average error is 0.03%. The maximum numerical error was observed for the low porosity case with the smallest roughness particles. The high magnitude of shear acting on the small roughness particles increased the numerical error in this case. We found that the ability of the realizable $k$-$\varepsilon$ model to predict the flow features is adequate for a comparative analysis between the flow around smooth and rough solid obstacles in the porous medium. It allows us to explore a wider range of parameter variations to identify the important features introduced by the surface roughness. Noting the shortcomings of the model, we recommend the use of DNS or Large Eddy Simulation (LES) in future studies for a more detailed analysis.

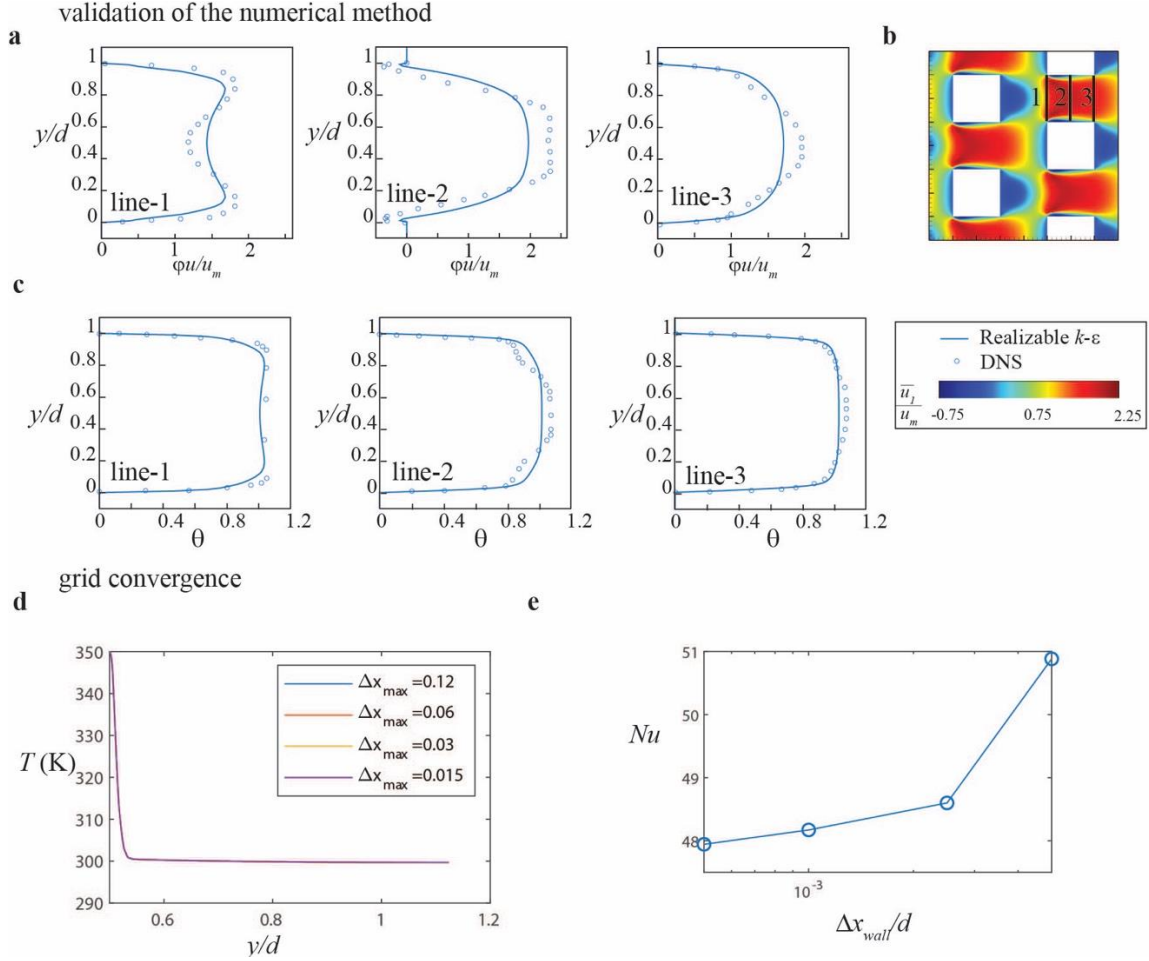

validation of the numerical method

grid convergence

**Fig. 2 a** Validation of the *x*- velocity profile at three locations inside the porous medium. **b** The *x*-velocity distribution inside the entire periodic porous medium. The three numbered black lines show the locations of the velocity and temperature profile plots. **c** Validation of the non-dimensional temperature (θ) profile at three locations inside the porous medium. The realizable *k*-ε model used in the present work is compared with direct numerical simulation results (Chu et al. 2019; Chu et al. 2018). **d** Temperature profiles along the *y*- axis predicted by grids are identical for different values of maximum grid size ($\Delta x_{max}$). **e** Nusselt number (*Nu*) variation with near-wall grid size ($\Delta x_{wall}$) shows less than 0.5% improvement in the prediction of *Nu* when $\Delta x_{wall}$ is less than 0.001.

Next, we performed a grid convergence study to decide the size of the computational cells in the domain. For this, we have chosen to simulate flow in porous media with a rough solid obstacle having $k_s = 0.005$, $w = 2k_s$, and φ = 0.8. This is the smallest roughness height and highest porosity we have used in section 3. The Reynolds number of the flow is 1000. We first varied the maximum grid size ($\Delta x_{max}$) while keeping the near-wall grid size ($\Delta x_{wall}$) at a constant value of 0.001. A near-wall grid size of 0.001 is necessary to ensure that the non-dimensional near-wall grid size based on the friction velocity ($\Delta y+$) is less than 1 in all the cases in section 3. We did not observe a clear convergent trend in the Nusselt number and the drag while varying $\Delta x_{max}$. The Nusselt number and the drag values predicted by all of the grids were within 1% of that predicted by the finest grid. We conclude that the maximum grid size is not a critical parameter in this study. We have plotted the temperature profiles for different values of $\Delta x_{max}$ to illustrate the independence of the temperature to the tested grid sizes (Fig. 2d). The near-wall grid size ($\Delta x_{wall}$) is more important in the present study. We have refined the near-wall grid until the point where the improvement in the Nusselt number prediction is less than 1% (Fig. 2e). Therefore, the choice of grid size for the subsequent simulations in the present work is $\Delta x_{max} = 0.03$ and $\Delta x_{wall} = 0.001$.

## 3 Results and Discussion

There are two length scales that characterize the surface roughness on the solid obstacle surface in this work – the surface roughness particle height and the surface roughness particle spacing (Fig. 1). Note that the aspect ratio of the roughness particles is 1 in all of the cases presented in this paper, and a study on the aspect ratio is outside the scope of the paper. The discussion has been divided into two sections focusing on the influence that the variation of the roughness particle height and spacing have on the surrounding flow. In each section, the porosity of the porous medium and the Reynolds number of the flow are chosen as the critical parameters since they alter the boundary layer surrounding the solid obstacle. The trends for the macroscale quantities – viscous drag, pressure drag, and surface-averaged Nusselt number, are reported. The microscale pressure, shear, and temperature distributions, as well as the flow streamlines are analyzed. They provide physical insight into the trends and a broader understanding of the contribution of the surface roughness particles to the microscale flow.

### 3.1 Surface roughness particle height

The surface roughness particle height ($k_s$) determines the minimum gap in between the surfaces of adjacent solid obstacles for a given porosity. Therefore, the surface roughness height has a significant influence on the microscale flow due to the flow constriction it introduces. In this case study, the height of the roughness particles has the following values: 0.005, 0.01, 0.05, and 0.1. The surface roughness particle spacing ($w$) is kept at a constant value of 2 times the roughness height. The $x$- pressure drag force ($F_{pressure}$) increases exponentially when the surface roughness particle height increases (Fig. 3a). The pressure drag force vector is calculated by integrating the product of periodic pressure and the normal vector to the surface area over the entire solid obstacle surface area. The viscous drag force vector is calculated by integrating the shear stress at the surface over the entire solid obstacle surface are. The forces are normalized by using $\rho u_m d$, whose value is equal to 1 in the present study. The pressure drag increase is more pronounced when the porosity decreases. The main source of the pressure drag is the flow stagnation that occurs at the leading vertices of the solid obstacle surface (yellow dots in Fig. 4). The roughness particles protrude into the void space in between the solid obstacles and constrain the fluid to flow through a smaller area causing an increase in the stagnation pressure when compared to a smooth solid obstacle. When the porosity decreases, the solid obstacle surfaces are brought close to one another. At high porosity, the flow more closely resembles other classic external flows such as the flow around a bluff body. At low porosity, the flow resembles classic internal flows such as the flow in channels. Therefore, the pore geometry is more important at low porosities, whereas the solid obstacle geometry is more important at high porosities. The roughness particles protrude deep into the pore space at low porosity and have a pronounced effect on the pressure drag.

In contrast to the $x$- pressure drag force, the $x$- viscous drag force ($F_{viscous}$) for the rough solid obstacle is less than that of the smooth solid obstacle for all of the reported cases (Fig. 3b). Note that the pressure drag force has a higher magnitude when compared to the viscous drag force due to the high Reynolds number of the flow. A recirculating vortex is formed behind each roughness particle due to flow separation at the vertex of each of the square roughness particles (Fig. 4 b and c). In the present case study where $w = 2k_s$, the recirculating vortices occupy the entire space in between the roughness particles. We have named the space in between the roughness particles as the roughness cavity (Fig. 1). The flow above the roughness particles is effectively "bridged" over the roughness cavity by the recirculating vortices. This behavior is consistent with the observations made for rough wall boundary layers with d- type roughness (Perry et al. 1969). By this mechanism, only 25% of the surface area of the top and bottom surfaces of the solid obstacle (defined in Fig. 1) is exposed to the high shear flow in the rough cases. Note that a coverage of 25% of the total solid obstacle surface area is equivalent to 50% of the projected surface area in the direction normal to the shear layer. The remaining surface area is covered by low shear flow arising from the recirculating vortices. In the smooth case, the entire top and bottom surfaces of the solid obstacle are exposed to high shear flow (Fig. 4a). Therefore, the boundary layer of the flow around the solid obstacle is modified by the roughness particles to decrease the velocity gradient in the surface normal direction, which results in a decrease in the viscous drag force. Even though the viscous drag reduction is experienced across all reported values of porosity, the increase in $k_s$ causes a decrease in $F_{viscous}$ for φ = 0.65 and 0.80, and an increase in $F_{viscous}$ for φ = 0.50. The large roughness particles introduce strong recirculating vortices that decrease the shear stress when compared to the small particles. At φ = 0.50, the roughness particles increase the center line velocity of the flow in the void space in between solid obstacles due to the confined geometry. This results in an increase in the velocity gradient, and consequently shear stress, for large roughness particles when compared to the small roughness particles at φ = 0.50.

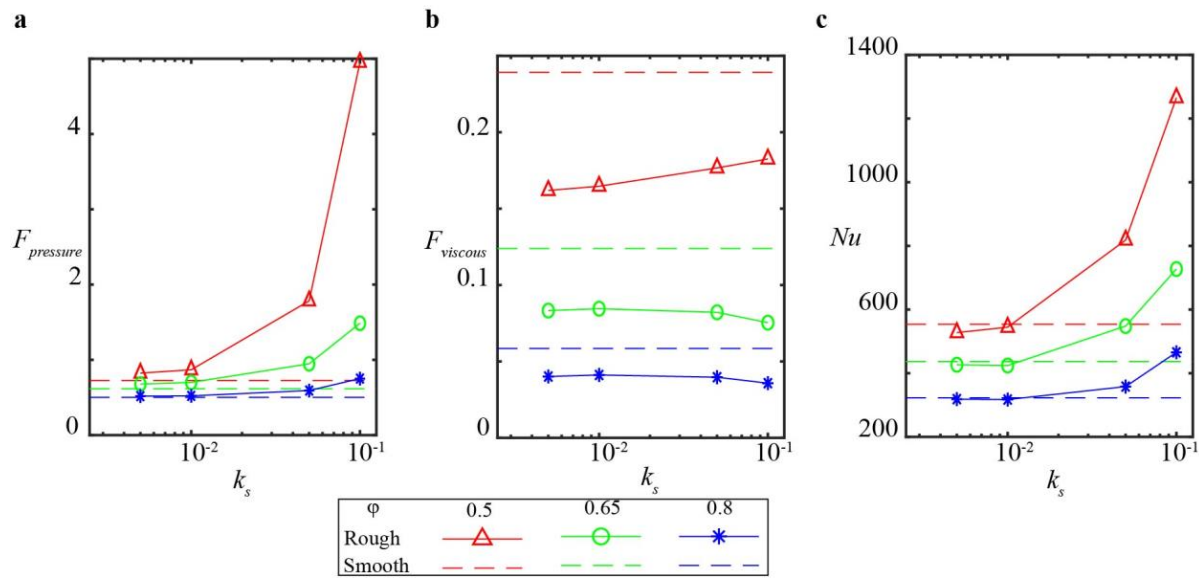

**Fig. 3** The variation of (**a**) $x$- pressure drag force, (**b**) $x$- viscous drag force, and (**c**) surface-averaged Nusselt number versus the surface roughness particle height.

To calculate the surface-averaged Nusselt number ($Nu$), the surface-averaged heat flux ($q_w''$) over the entire solid obstacle surface is used along with the perimeter of the solid obstacle surface as the characteristic length scale (equation 18). The diameter of the solid obstacle is more commonly used as the characteristic length scale, but it poses a problem in the case of rough solid obstacles. Rough solid obstacles have a greater solid obstacle surface area than the smooth solid obstacles when the area contributed by the roughness particles is considered. This causes a jump in the surface-averaged Nusselt number between the smooth and rough cases even in the limiting case where the rough solid obstacle is virtually smooth. To overcome this, the additional area contributed by the roughness particles is included in the characteristic length scale for the surface-averaged Nusselt number calculation by using the perimeter of the solid obstacle. This produces a continuous trend for the surface-averaged Nusselt number with the roughness height in Fig. 3c. The surface-averaged Nusselt number would otherwise decline by a factor of 2 from the smooth case to the rough ($k_s = 0.005$) case even though the total heat transfer rates are similar. The Nusselt number is thus defined as:

$$Nu = \frac{4q_w^{''}(d+2n_p k_s)}{(T_w - T_{in})\lambda} \tag{18}$$

When the roughness particle height ($k_s$) is increased, there is a cross-over from a heat transfer reduction to a heat transfer enhancement due to surface roughness when compared to the smooth case. Comparing Fig. 4 c to b, the large roughness particles ($k_s = 0.1$) protrude deeper into the thermal boundary layer when compared to the small roughness particles ($k_s = 0.005$). In the case of $k_s = 0.1$, the recirculation vortex entrains colder fluid from the incoming flow into the roughness cavity resulting in increased convection heat transfer from the solid obstacle surface to the fluid. However, the heat flux inside the roughness cavity for the rough solid obstacle ($k_s = 0.1$) is still less than the heat flux from the surface of the smooth solid obstacle. This is due to the low temperature gradient inside the roughness cavity caused by flow recirculation (Fig. 4c). The deficit in heat flux magnitude inside the roughness cavity is overcome by the higher surface area of the rough solid obstacle compared to the smooth solid obstacle. The result is an overall increase in the total heat transfer rate and the surface-averaged Nusselt number for the solid obstacle in the rough case with $k_s = 0.1$ compared to the smooth case. The heat transfer enhancement due to roughness at $k_s = 0.1$ is higher when the porosity is decreased. At low porosity, the heat flux from the roughness particle and the roughness cavity to the surrounding fluid is increased by the increase in shear. The increase in the shear causes a decrease in the width of the thermal boundary layer, which in turn increases the heat flux. Note that the heat transfer enhancement for large roughness particles comes with the cost of exponential increase in drag (Fig. 3a).

In the case of $k_s = 0.005$, the roughness particles do not entrain a significant amount of cold fluid from the incoming flow. The roughness cavity in this case is filled with stagnant, heated fluid whose temperature is close to the solid obstacle wall temperature. Since the roughness particle height of $k_s = 0.005$ is less than the width of the thermal boundary layer at the solid obstacle surface, the low heat flux inside the roughness cavity only decreases the Nusselt number by ~1.5% at $\varphi = 0.80$ and ~5% at $\varphi = 0.50$. The size of the recirculating vortex and the roughness cavity is only 0.5% of the side length of the square solid obstacle. Therefore, the temperature gradient at the solid obstacle surface is not significantly modified by the small roughness particles ($k_s = 0.005$) when the temperature distributions are compared between the smooth and rough cases (Fig. 4 a and b).

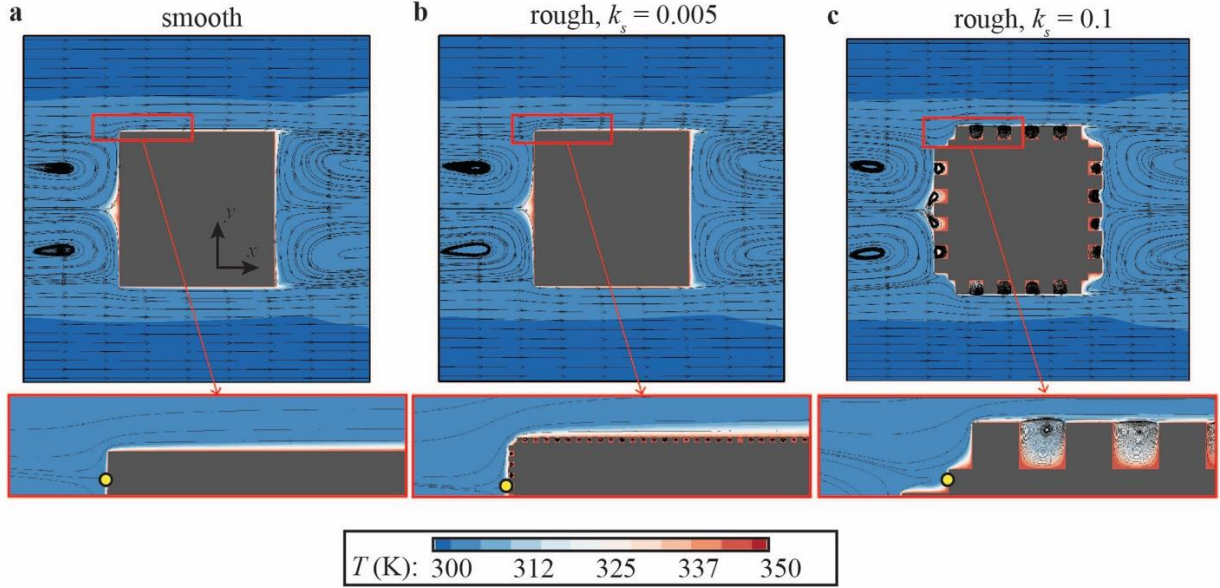

**Fig. 4** Microscale flow streamlines plotted on top of the temperature distribution for the (**a**) smooth, (**b**) $k_s = 0.005$, and (**c**) $k_s = 0.1$ cases ($w = 2k_s$, $\varphi = 0.8$, $Re = 1000$). The yellow dots show the flow stagnation points.

When the Reynolds number of the flow is increased from $Re = 1000$, the qualitative observations made previously remain valid at $Re = 5000$ and $10000$. The pressure and viscous drag forces decrease with an increase in Reynolds number (Fig. 5 a and b). To increase the Reynolds number in the present work, the viscosity of the fluid is decreased while keeping $d$, $u_m$, and $\rho$ at a constant value of 1 so that the governing equations for the conservation of momentum are non-dimensional (equations 3-8). Therefore, an increase in the Reynolds number will result in the decrease of the pressure and viscous forces on the solid obstacle. The surface-averaged Nusselt number increases with the Reynolds number due to the decrease in the width of the thermal boundary layer brought by the flow of less viscous fluid over the solid obstacles. The increase in the surface-averaged Nusselt number with the increase Reynolds number is observed for both the rough and smooth cases. However, the increase in the Reynolds number amplifies the influence of surface roughness on the drag force and the heat transfer rate. The increase in the Reynolds number increases the percentage of total drag reduction for the rough $k_s = 0.005$ case from 0.5% at $Re = 1000$ to 2.7% at $Re = 10000$ when compared to the respective smooth cases. The drag increase for the rough $k_s = 0.1$ case changes from 40% at $Re = 1000$ to 80% at $Re = 10000$ when compared to the respective smooth cases. The heat transfer reduction for the rough $k_s = 0.005$ case changes from 1.2% at $Re = 1000$ to 6% at $Re = 10000$ when compared to the respective smooth cases. The heat transfer enhancement for the rough $k_s = 0.1$ case changes from 44% at $Re = 1000$ to 80% at $Re = 10000$ when compared to the respective smooth cases.

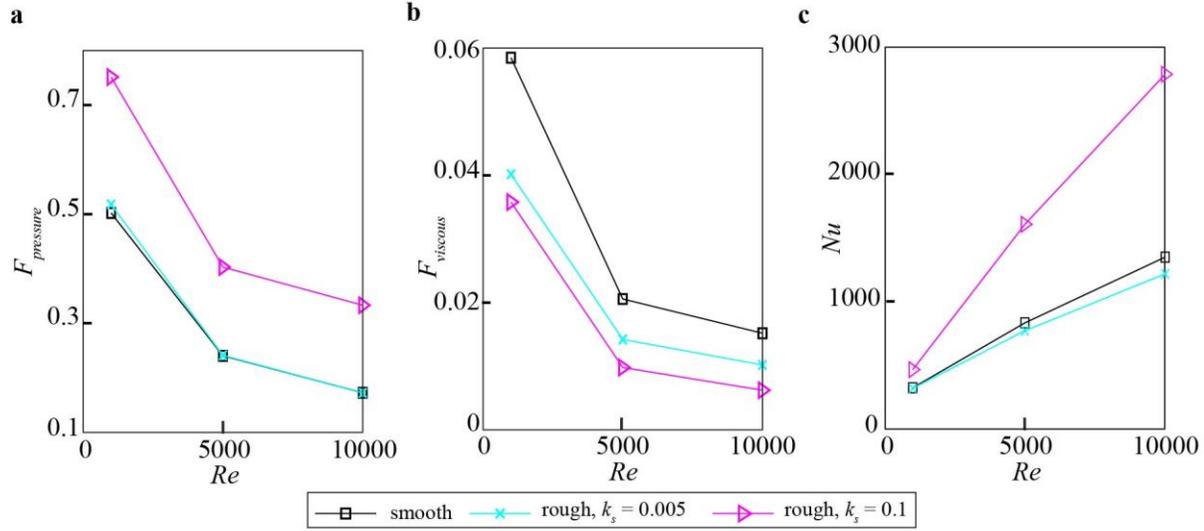

**Fig. 5** The variation of the (**a**) pressure force, (**b**) viscous force, and (**c**) Nusselt number with the Reynolds number for the smooth and rough ($k_s = 0.005$ and $k_s = 0.1$) cases ($w = 2k_s$, $\varphi = 0.8$).

When the Reynolds number is increased, the width of the thermal boundary layer in the normal direction of the solid obstacle surface decreases (Fig. 6). In the smooth case, the width of the thermal boundary layer consistently decreases with the decrease in the fluid viscosity. Whereas in the rough cases, the roughness particles modify the heat transfer from the solid obstacle surface depending on the roughness particle height. In the rough $k_s = 0.005$ case, the roughness particle height limits the decrease in the width of the thermal boundary layer. Flow recirculation inside the roughness cavity insulates the solid obstacle surface from the surrounding flow at all of the Reynolds numbers tested. Above the roughness layer (Fig. 1), the decrease in the thermal boundary layer width is similar to that of the smooth case (Fig. 6a). Thus, the effective boundary layer width is therefore always greater than the roughness particle height ($k_s = 0.005$) due to the flow recirculation inside the roughness cavity. In the rough $k_s = 0.1$ case, a thermal boundary layer develops inside the cavity in between the roughness particles due to the entrainment of the cold fluid from the incoming flow by the recirculating vortex. The heat transfer rate in this case is not limited by the roughness particle height since the flow in the cavity interacts with the flow above the roughness particles leading to mixing. The temperature gradient at the surface inside the roughness cavity increases when the Reynolds number is increased (Fig. 6b). The increase in the temperature gradient increases the heat flux that is removed from the surface through the roughness cavity.

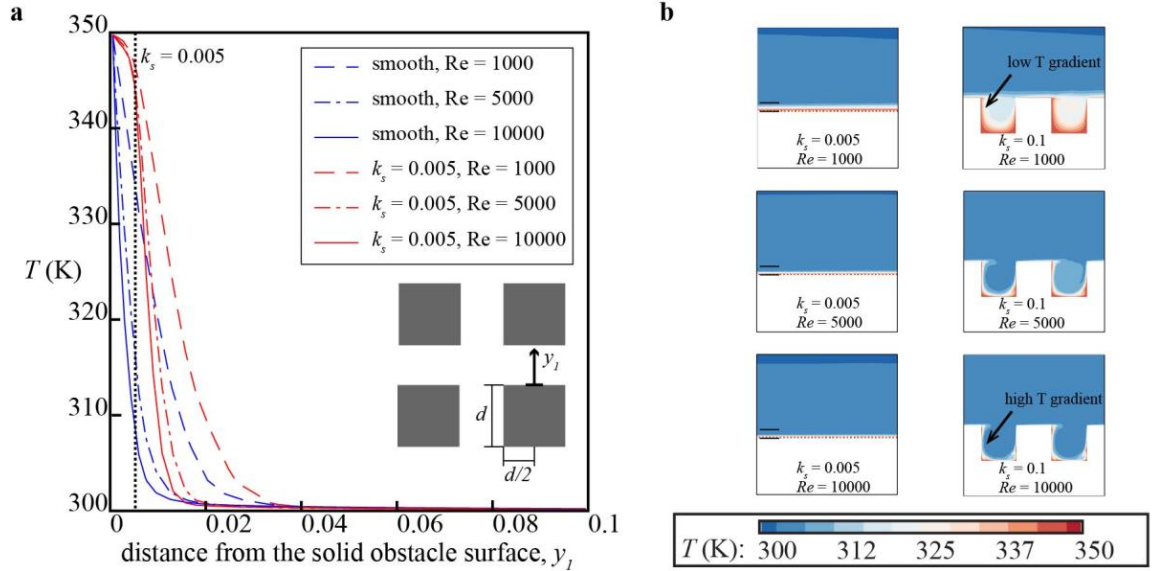

**Fig. 6** The change in the width of the thermal boundary layer due to surface roughness with the increase in Reynolds number shown for (**a**) $k_s = 0.005$ for a vertical line at the top face of the solid obstacle surface and (**b**) $k_s = 0.005$ and $k_s = 0.1$ in the cavity in between roughness particles ($w = 2k_s$, $\varphi = 0.8$).

Following these observations, there are two regimes of the surface roughness particle height that uniquely influence the surrounding flow. In the fine roughness regime, the roughness particles with heights $k_s = 0.005$ and $k_s = 0.01$ are small and numerous on the solid obstacle surface. The roughness particles do not drastically alter the microscale flow around the solid obstacle when compared to the smooth case. The roughness particles are located deep inside the boundary layer around the solid obstacle such that the recirculating vortex that forms in the roughness cavity is virtually stationary with low rotational velocity. The roughness particles in the fine roughness regime cause the increase in the boundary layer width on the solid obstacle surface. The fine roughness regime is observed in Fig. 4b. The flow patterns in the porous medium for the rough solid obstacle are similar to those observed for the smooth solid obstacle (Fig. 4a). In the coarse roughness regime, the roughness particles with heights $k_s = 0.05$ and $k_s = 0.1$ are large enough that they modify the geometry of the solid obstacle. The influence of the roughness particles extend beyond the boundary layer surrounding the solid obstacle. Flow mixing is enhanced near the solid obstacle surface due to the formation of strong recirculating vortices in the roughness cavity. The large roughness particles also obstruct the flow leading to a drastic increase in the drag force in this regime. The roughness particles in the coarse roughness regime cause the change of the flow patterns surrounding the solid obstacle. The coarse roughness regime is observed in Fig. 4c where the roughness particles cause strong flow recirculation and mixing in the roughness layer.

### 3.2    Surface roughness particle spacing

Solid obstacle surface roughness leads to the formation of recirculating vortices behind the roughness particles, which bridge the flow over the cavity in between two adjacent roughness particles for $w = 2k_s$ as shown in section 3.1. The close proximity of the adjacent roughness particles when $w = 2k_s$ ensures that the recirculating vortex completely occupies the cavity, thereby decreasing heat transfer in the fine roughness regime. Evidently, the surface roughness particle spacing is a crucial parameter for the heat transfer from the solid obstacle surface inside the cavity to the surrounding flow. In this case study, the roughness particle spacing ($w$) has been set to the following values: $2k_s$, $4k_s$, $8.33k_s$, and $16.67k_s$. The surface roughness particle height is fixed at a constant value of $k_s = 0.01$. Since the solid obstacle surface area is finite in this case, the increase in the roughness particle spacing will lead to a decrease in the number of roughness particles on the solid obstacle surface. The number of roughness particles along a single side of the square solid obstacle corresponding to the roughness particle spacing has been set to the following values: 50, 25, 12, and 6, respectively. This is why the case study is performed for the fine roughness regime. Changing the roughness particle spacing in the coarse roughness regime will decrease the number of roughness particles on the solid obstacle surface such that the resulting solid obstacle cannot be considered rough.

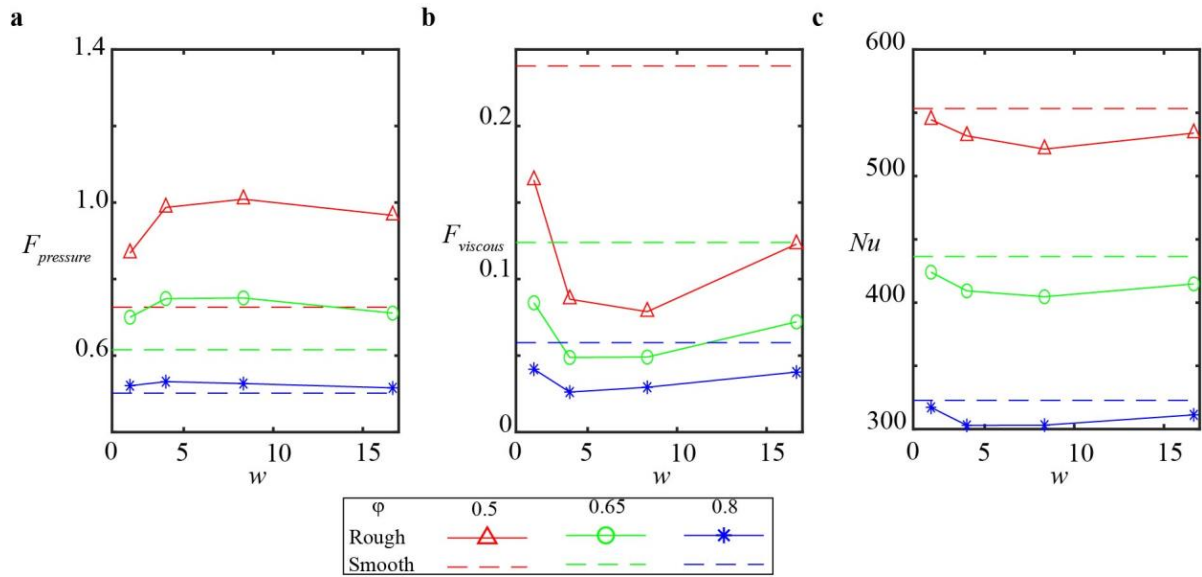

**Fig. 7** The variation of (**a**) $x$- pressure drag force, (**b**) $x$- viscous drag force, and (**c**) surface-averaged Nusselt number versus the surface roughness particle spacing ($k_s = 0.01$, $Re = 1000$).

The fundamental observations about the increase in pressure drag and the decrease in viscous drag and surface-averaged Nusselt number due to roughness follow from section 3.1. The roughness particle spacing primarily influences the flow inside the roughness layer on the solid obstacle surface. The dependence of the pressure drag, viscous drag, and the surface-averaged Nusselt number on the roughness particle spacing is non-monotonic (Fig. 7), unlike the dependence on the roughness particle height. Consider the variation of roughness particle spacing at $\varphi = 0.80$. When the roughness particle spacing is increased, there are two competing factors – the change in the flow inside the cavity between the roughness particles and the decrease in the number of roughness particles. The size of the recirculating vortex that is formed behind the roughness particle scales with the roughness particle height. For $w = 2k_s$, a single recirculating vortex occupies the entire cavity as described in section 3.1. The flow streamlines above the roughness layer are virtually undisturbed as it is bridged over the recirculating vortex in the roughness cavity (Fig. 8a). This is similar to the volumetric sheltering effect reported in Yang et al. (2016). For $w = 4k_s$, the roughness particles are still spaced close enough that a pair of recirculating vortices are formed in the cavity instead of the single vortex at $w = 2k_s$. The flow streamlines over the roughness layer become tortuous when they enter the roughness cavity due to the increase in $w$. The recirculating vortex pair does not completely occupy the roughness cavity in this case, even though the flow inside the cavity does not reattach (Fig. 8b).

The pressure drag increases with the increase in the roughness particle spacing from $w = 2k_s$ to $4k_s$ (Fig. 7a). To determine the source of the increased drag force that arises when the roughness spacing is increased from $2k_s$ to $4k_s$, the magnitude of pressure has to be standardized across the different cases. The assumption of periodic pressure distribution in the REV implies that the magnitude of pressure is not fixed at a reference value. The selection of a reference pressure location inside the REV is not straightforward since several flow patterns emerge in the porous medium such as flow recirculation, and flow acceleration in adverse and favorable pressure gradients. In the present work, the reference pressure used to calculate a standardized pressure ($P_{std}$) is set equal to the periodic pressure at the midpoint in between 4 neighboring solid obstacles at a location ($\pm s/2, \pm s/2$) from the centroid of any solid obstacle. When $w$ is increased from $2k_s$ to $4k_s$, the sum of the standardized pressure drag forces on the front and rear faces of the solid obstacle (defined in Fig. 1) decreases by 0.2%. Therefore, the increase in the pressure drag is not caused by the flow stagnation pressure on the front face of the solid obstacle or the flow separation on the rear face of the solid obstacle. The standardized pressure drag on the top and bottom faces increases by 45% from $w = 2k_s$ to $4k_s$. The increase in the pressure drag arises due to the change in the recirculating vortex system in the roughness cavity. The formation of a recirculating vortex pair for the $w = 4k_s$ case increases the tortuosity of the flow streamlines between the roughness layer and the flow above it when compared to the $w = 2k_s$ case (Fig. 8b). Increased flow streamline tortuosity increases the stagnation pressure on the surface of the roughness particle. Higher peak pressure is observed in the front face of the roughness particle for the $w = 4k_s$ case when compared to the $w = 2k_s$ case (compare the distribution of $P_{std}$ in Fig. 8b to Fig. 8a). Even though the peak pressure acts on the small surface area of the roughness particle face, the roughness particles are numerous and the increase in the pressure drag is compounded.

The formation of the recirculating vortex pair also causes the viscous drag force to decrease from $w = 2k_s$ to $w = 4k_s$. When $w$ increases from $2k_s$ to $4k_s$, the width of the cavity increases along with the fraction of the solid obstacle surface area that is covered by recirculating flow. In the smooth case, the entire top and bottom faces of the solid obstacle are subjected to strong shear flow. In the $w = 2k_s$ case, only 25% of the surface area of the top and bottom faces of the solid obstacle is exposed to the strong shear flow, while the remaining surface area is covered by the recirculating vortex. In the $w = 4k_s$ case, the recirculating vortices cover 83% of the surface area of the top and bottom faces of the solid obstacle, which decreases the viscous drag force substantially. Since the recirculating flow has low shear at the wall when compared to the attached flow, the viscous drag decreases according to the fraction of the solid obstacle area covered by recirculating flow. Note that the viscous drags on the front and rear faces of the solid obstacle are 3 orders of magnitude smaller than those on the top and bottom faces. The recirculating flow also decreases the heat flux from the solid obstacle surface when compared to attached flow. Therefore, the surface-averaged Nusselt number decreases when $w$ is increased from $2k_s$ to $4k_s$ due to the formation of a pair of recirculating vortices in the roughness cavity.

For $w = 8.33k_s$ and $16.67k_s$, recirculating vortices do not occupy the entire roughness cavity (Fig. 8 c and d). Behind each roughness particle, the formation of the recirculating vortex is followed by flow reattachment inside the roughness cavity. This leads to attached flow over the remaining width of the cavity. The attached flow inside the roughness cavity forms a flow stagnation when it approaches the neighboring roughness particle resulting in high pressure in that region.

Therefore, the flow inside the roughness cavity for high roughness particle spacing consists of flow separation and recirculation, flow reattachment, and flow stagnation (called "tripping" hereafter). The flow tripping results in the increase of the pressure drag force on the individual roughness particles due to the increase in the stagnation pressure acting on the roughness particles. This behavior is consistent with the observations made for rough wall boundary layers with k- type roughness (Perry et al. 1969). However, the pressure drag force on the entire solid obstacle decreases due to the decrease in the number of roughness particles on the obstacle surface. The viscous drag force and the Nusselt number increase from $w = 4k_s$ to $16.67k_s$ due to the decrease in the surface area on the solid obstacle that is covered by flow recirculation. The flow reattachment during the flow tripping process increases the shear stress and the heat flux in the large attached flow regions of the cavity.

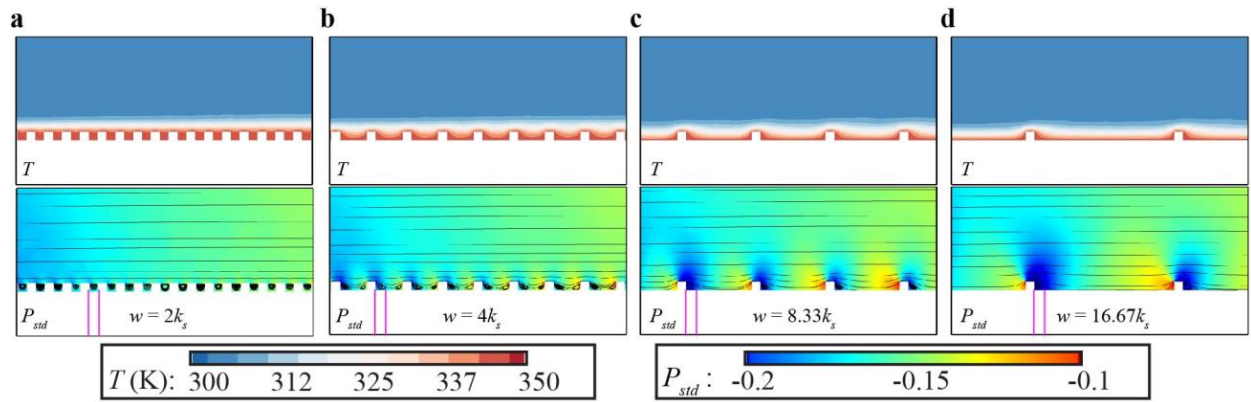

**Fig. 8** Temperature and standardized pressure distribution for (**a**) $w = 2k_s$, (**b**) $w = 4k_s$, (**c**) $w = 8.33k_s$, and (**d**) $w = 16.67k_s$ ($\varphi = 0.8$, $Re = 1000$, $k_s = 0.01$). Flow streamlines are plotted on top of the pressure distribution. We drew two pink vertical lines that are separated by $k_s$ distance to show the similar size of the recirculating vortex.

When the porosity of the porous medium decreases, the shear stress that is acting on the top and bottom faces of the solid obstacle surface increases. In the roughness layer, the transition from a single recirculating vortex to a pair of recirculating vortices to flow tripping is delayed when the porosity is decreased. For $\varphi = 0.50$, a single recirculating vortex is formed behind the roughness particle in the $w = 4k_s$, which continues to bridge the flow over the roughness layer. In this case, the aspect ratio of the recirculating vortex is such that the vortex height is equal to the roughness particle height but the vortex width is equal to the width of the roughness cavity. The change in the vortex aspect ratio is caused by the increase in the flow anisotropy due to the increase in the shear stress. The increase in the vortex width is observed in the $w = 8.33k_s$ and $16.67k_s$ cases as well. As the porosity decreases, the shear stress, and consequently the width of the recirculating vortex in the roughness cavity, increases (Fig. 9). The delay in the transition of the recirculating vortex causes the shift in the trends for the pressure and viscous drag and the surface-averaged Nusselt number in Fig. 7.

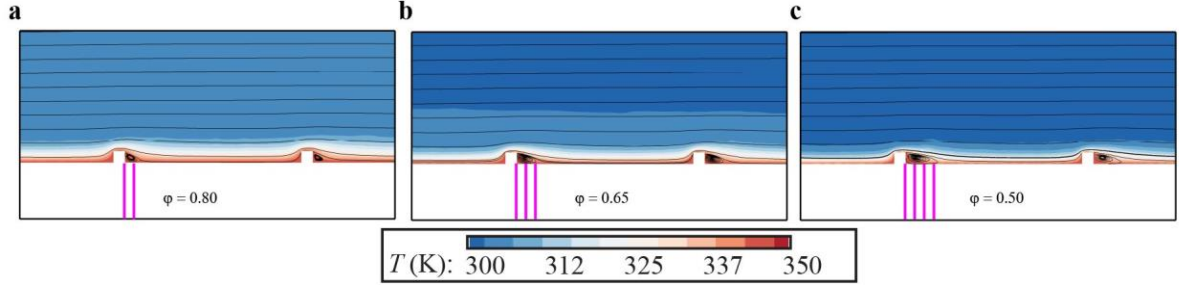

**Fig. 9** Flow streamlines overlaid on the temperature distribution for different values of porosity ($Re = 1000$, $w = 16.67k_s$, $k_s = 0.01$). We drew pink vertical lines that are spaced $k_s$ distance apart to show the increase in the width of the recirculating vortex when porosity decreases.

When the Reynolds number of the flow increases, the pressure and viscous drag forces decrease, which is similar to the effect of the Reynolds number on the roughness height (section 3.1). The decrease in the viscosity of the fluid leads to a decrease in the viscous drag as well as the pressure drag since there is less resistance. For $w = 8.33k_s$, the increase in the Reynolds number causes the Nusselt number to increase such that it transitions from a heat transfer decline at $Re = 1000$ to a heat transfer enhancement at $Re = 10000$ when compared to the smooth case (Fig. 10a). The heat flux at the surface of the roughness cavity is 20% less at $Re = 1000$ and 6% less at $Re = 10000$ than the heat flux at the same location for a smooth obstacle surface. When the Reynolds number increases, the width of the thermal boundary layer inside the roughness cavity decreases, which causes the local heat flux in the cavity to increase. Note that the width of the thermal boundary layer is not limited by the roughness particle height for $w = 8.33k_s$, unlike for $w = 2k_s$ (Fig. 10b). This is because flow tripping occurs in the roughness cavity for $w = 8.33k_s$ instead of the flow recirculation for $w = 2k_s$.

The cross-over from heat transfer reduction to enhancement on the solid obstacle surface when the Reynolds number increases (Fig. 10a) is due to the increase in the heat flux from the roughness particles to the flow over it. At the location of flow stagnation on the roughness particle, the local heat flux is 2.6 times higher at $Re = 1000$ and 5 times higher at $Re = 10000$ than for the smooth case at a similar location. Even though the roughness particles are small, their effect on heat transfer is compounded over the entire solid obstacle surface leading to a heat transfer enhancement at higher Reynolds numbers. Therefore, the primary factors for heat transfer enhancement with an increase in the Reynolds number are – (1) the decrease in the thermal boundary layer width over the entire solid obstacle surface and (2) the increase in the heat transfer rate from the roughness particle at the flow stagnation location.

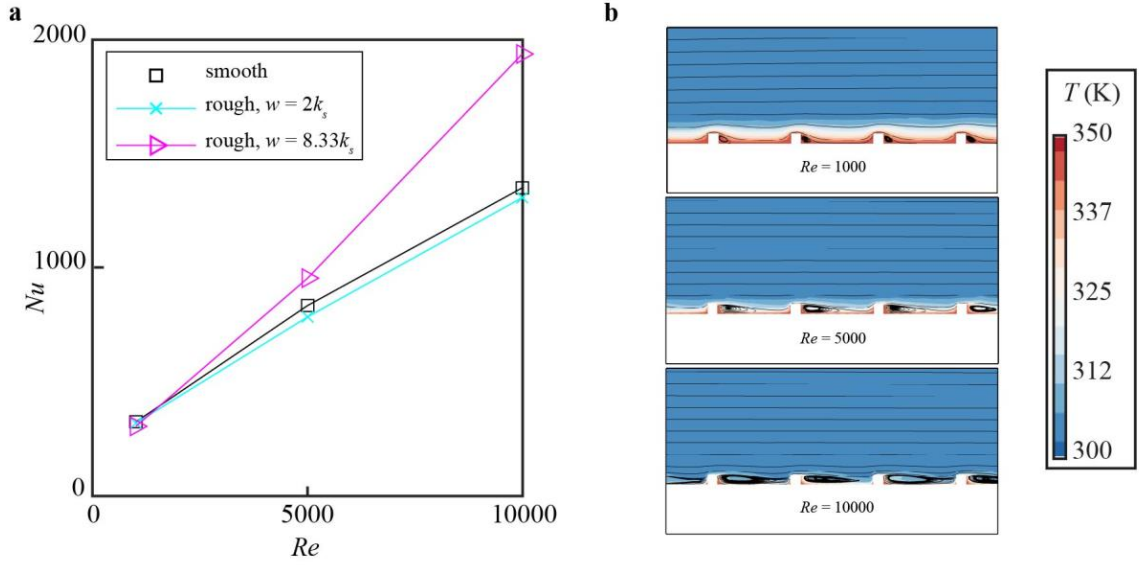

**Fig. 10 a** The variation of surface averaged Nusselt number with Reynolds number for smooth and rough solid obstacles with different roughness particle spacing. **b** The temperature distribution around the rough solid obstacle ($\varphi = 0.8$, $Re = 1000$, $w = 8.33k_s$) at different Reynolds numbers.

## 4    Summary

Solid obstacle surface roughness is an important factor in the design of porous media for turbulent convection flows. Recirculating vortices are formed behind each of the roughness particles on the surface of the solid obstacle. The characteristics of the recirculating vortices and the flow inside the roughness cavity are dependent on two geometric parameters of the roughness particles – roughness particle height and spacing. Other important parameters identified in this work are the porosity of the porous medium and the Reynolds number of the flow. We define two flow regimes for porous media with rough solid obstacles based on the roughness height. The fine roughness regime ($k_s = 0.005d$ and $0.01d$) is characterized by microscale flow patterns that are virtually identical to those for a smooth solid obstacle. The roughness particles decrease the temperature and velocity gradient normal to the solid obstacle surface, but they do not influence the flow at the center of the pore. The roughness particles only produce a marginal change in the macroscale quantities such as the drag and the surface-averaged Nusselt number when compared to a smooth solid obstacle. The coarse roughness regime ($k_s = 0.05d$ and $0.1d$) occurs when large roughness particles that modify the geometry of the solid obstacle are present on the solid obstacle surface. This in turn produces microscale flow patterns that are different from the smooth case. The influence of the roughness particles extends to the entire flow field inside the porous medium.

In the fine roughness regime, the pressure drag increases due to the increase in the stagnation pressure acting on the solid obstacle surface. The viscous drag decreases due to the decrease in shear stress caused by flow separation during the formation of the recirculating vortices behind the roughness particles. In the case of $k_s = 0.005d$, there is a 0.5% reduction in the total drag that is acting on the solid obstacle compared to the smooth case at $Re = 1000$. The viscous drag reduction in this case is greater than the pressure drag increase. The surface-averaged Nusselt number in the fine roughness regime is always less than that of the smooth case. Even though surface roughness increases the surface area of the solid obstacle, the roughness cavity is insulated by the recirculating vortex leading to heat transfer reduction. The roughness particles in the fine roughness regime are small enough that the rough solid obstacle behaves like a smooth solid obstacle with a thin layer of insulating fluid surrounding it. In the coarse roughness regime, the roughness particles protrude into the pore space and constrict the flow in between the solid obstacles resulting in a multifold increase in the pressure drag. The viscous drag for the rough solid obstacle continues to be less than that of the smooth solid obstacle, but the total drag is substantially higher for the rough solid obstacle due to the large magnitude of the pressure drag. In the coarse roughness regime, the surface-averaged Nusselt number is higher than that for the smooth case, resulting in heat transfer enhancement at the cost of increased drag. The large roughness particles in the coarse roughness regime entrain cold fluid from above the roughness layer, which increases the heat flux at the roughness cavity when compared that of the fine roughness regime. The combined effect of flow entrainment and the increased surface area of the solid obstacle due to roughness leads to heat transfer enhancement in the coarse roughness regime.

Two types of surface roughness are defined based on the roughness particle spacing. When the roughness particles are close to one another ($w = 2k_s$ and $4k_s$), the roughness cavity is occupied by recirculating vortices that bridge the flow over the roughness particles without flow reattachment inside the cavity. When the roughness particles are far apart ($w = 8.33k_s$ and $16.67k_s$), the flow inside the roughness cavity reattaches downstream of the recirculating vortex and stagnates on the front face of the neighboring roughness particle (called flow tripping). The roughness spacing determines the surface area of the roughness cavity that is occupied by recirculating flow, reattached flow, and flow stagnation. The recirculating vortices decrease the heat flux at the roughness cavity, whereas the reattached and stagnation flows increase the heat flux at the roughness cavity. The heat transfer reduction at $w = 2k_s$ transitions to heat transfer enhancement at $w = 16.67k_s$ since the fraction of the surface area of the roughness cavity occupied by the recirculating vortex decreases. However, the increase in the roughness spacing also decreases the total number of roughness particles on the solid obstacle surface. This causes the flow behavior around the rough obstacle to approach that of the smooth case when the roughness particle spacing is large. Therefore, the influence of the roughness particle spacing is governed by two factors – flow tripping in the roughness layer and the number of roughness particles on the solid obstacle surface.

The heat transfer from the rough solid obstacles is directly related to the width of the thermal boundary layer around the solid obstacle surface. When the porosity of the porous medium is decreased, the solid obstacles surfaces are brought closer to one another. The close proximity of the solid obstacle surfaces at low porosity amplifies the influence of the roughness particles. The thermal boundary layer width is also decreased when the Reynolds number increases. When the porosity decreases or the Reynolds number increases, the shear stress acting on the flow in between the solid obstacles at the edge of the roughness layer increases. The reduction in the thermal boundary layer width in high shear flow increases the surface averaged Nusselt number.

To improve the heat transfer rate from rough solid obstacles, the roughness particles should be fine and spaced far apart to minimize the detrimental effects of the recirculating vortices. Large roughness particles can improve heat transfer at the cost of high drag on the solid obstacles, which would require a higher capacity pump to overcome. Total drag reduction can be achieved for porous media using fine roughness particles on the solid obstacle surface by taking advantage of the viscous drag reduction caused by the recirculating vortices. Total drag reduction was only observed in the high porosity cases. For the low porosity cases, the total drag reduction is not observed due to the substantial increase in the pressure drag, which overshadows the viscous drag reduction.

The present work focuses on the use of RANS flow solutions in a 2D computational domain for square roughness particles. There is a lot of scope for future work since the flow field is strongly dependent on the microscale solid obstacle geometry. The variation of the aspect ratio, the dimensionality, and the shape of the roughness particle will provide further insight. For example, geometries that allow the shift in the location of the flow separation may have a different flow behavior. In our study, we identified that the surface roughness has greater influences on the drag and Nusselt number at lower porosities. Future work should investigate whether surface roughness is relevant at high porosity or if the solid obstacle shape is the only consideration. Scale resolving studies using DNS or LES will provide valuable information about the length and time scales of turbulence introduced by roughness, as well as highlight the presence of transient phenomena that the RANS study cannot resolve.

**Acknowledgments** AVK acknowledges the support of the National Science Foundation (award CBET-2042834) and the Alexander Humboldt Foundation Research Award.

**Declarations**

**Conflict of Interest** The authors declare that they have no conflict of interest.